# Efficient, Single Hop Time Synchronization Protocol for Randomly-Connected WSNs

Ali Al-Shaikhi, *member, IEEE* and Ahmad Masoud *Member, IEEE*

the Electrical Engineering Department, King Fahad University of Petroleum & Minerals, Dhahran, 31261, Saudi Arabia (e-mail: shaikhi@kfupm.edu.sa; masoud@kfupm.edu.sa)

*Abstract*—This paper develops a fast, accurate and energy-efficient time synchronization protocol in wireless sensor networks that operate in harsh environments. The suggested protocol concept is based on an electrical physical metaphor. The protocol treats the nodes times as the states of an unconditionally stable discrete dynamical system that uses only single hop communication. The system can terminate at a low number of message exchanges while still in the transient phase. It can yield high quality time estimates with accuracies that are fractions of the clock at which node communication takes place. The synchronizer is developed and its capabilities are demonstrated by simulation and physical experiments on the MICA-Z wireless sensor network

## I. INTRODUCTION

Wireless sensor networks (WSNs) consist of multiple devices called sensor nodes that are spread over a geographical area [1],[2]. Each of these sensor nodes consists of transducers or sensors, radio transceiver with wireless capabilities, low complexity processing units, and power supply. These systems are usually made of cheap components and deployed in harsh environments where communication is limited and the probability of node failure is high. Under such conditions, the nodes of the network can only communicate in an opportunistic manner making multi-hop communication and/or a fixed communication topology unattainable. The nodes have to reduce their message exchange rate to the minimum possible since communication can quickly drain the node battery. Despite the above and other difficulties, a WSN is expected to carry-out its function in a timely and practical manner.

Making all the WSN nodes agree on a common time which is most likely that of a gateway node is one of the activities (e.g. routing, topology/geometry reconstruction, etc.) the network has to perform in order to function. While the literature abounds with techniques on synchronization [3]-[5], most of these methods do not suit operation in harsh environments. They assume that the WSN has a specific stationary connectivity, node labeling or multi-hop communication, among other assumptions.

A promising approach that has the potential to address the requirements of operating in harsh environments uses consensus control [6] to synchronize the nodes of a WSN. In this approach, a dynamical system, which uses the local times of its nearest neighbors as input, acts as a virtual clock of a node [7,8]. However, there are serious issues that still need to be addressed before this approach becomes practical for use with WSNs deployed in harsh environments. One of the issues with this approach is that it uses too many communications. This can quickly deplete the nodes' batteries. Also, the accuracy and rate of convergence from existing methods in this approach may not be practical. Stability may depend on the type of topology the network has or on this topology remaining stationary during operation.

In this paper, we suggest a dynamical system that can operate as a virtual clock individual nodes of a WSN may use to estimate the gateway node (GWN) time after which they may use the estimate to reset their physical clocks. The system accepts, as input, the values neighboring nodes have for their local time and produces, as output, an estimate of the GWN time. It does that by averaging the input local times of the neighboring nodes. The system can terminate its local time update while in the transient phase by using a stopping criterion derived from the dynamical nature of the estimate time series. The nodes constituting the WSN are guaranteed to converge each to a local time that is close to that of the GWN provided that a path exist from the GWN that spans all the nodes of the WSN.

## II. THE SYNCHRONIZATION PROCEDURE

In this section, the analogy to an electric circuit used to develop the synchronizer is explained. Also, the components of the system are presented.

*A. The Resistive Network Metaphor*

To have a reasonable ability to function in harsh environments, the correctness of a node behavior (micro-properties of the WSN) should not be conditioned on global properties the WSN must possess. For example, a manner in which a node function that assumes a certain network connectivity or a labeling system is highly unlikely to succeed in harsh environments where the connectivity and labeling schemes can be easily invalidated by a node getting damaged. The manner in which a node function should be indifferent to the label of a node if the node is part of the group constituting the WSN. This immediately excludes the possibility of a node using multi-hop communication. This indifference also undermines a node's ability to make use of a global connectivity scheme the network may have.

It may appear difficult to find a node-level synchronization procedure that requires only single hop communication, is not strongly reliant on a communication (connectivity) arrangement and provides guarantees on the networks ability to function. However, a quick examination of electrical networks, as shown in Figure-1, proves otherwise [9]. If the voltage of any node in an electrical network is fixed to a certain value V, all the other non-isolated nodes (i.e a path exists from a node to the fixed voltage node) will register the same voltage. The nodes of the electric network function in a manner that is indifferent to the labels of the other nodes or their connectivity scheme provided that a path from the fixed voltage node exists that spans all the nodes of the network. Moreover, deleting or inserting new nodes into the network will not affect the operation if the new nodes are part of a spanning path or the removed nodes do not render the set of spanning paths the empty set. As can be seen, a strong analogy exists between the local time of a WSN node and the voltage of a resistive electrical circuit. This analogy is exploited to build a WSN synchronizer with the desired properties.

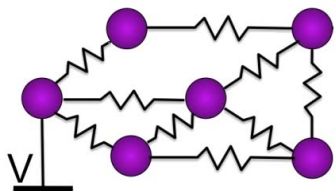

Figure-1: Analogy between a resistive network and a WSN

B. Node Local Time Update

Based on the electric analogy and assuming that we start at a reference update time at iteration $n$ of the GWN, $t_{N+1}$, this time is equal to the time of message exchange ($\Delta T$) multiplied by the iteration value $n$ ($t_{N+1}(n) = \Delta T \cdot n$). The updated time equation for each node $i$, $t_i$, is given by:

$$t_i(n+1) = \frac{1}{C_i} \sum_{j \in \chi_i} t_j(n) \quad j = 1,..N \qquad (1)$$

where N is the number of nodes in the WSN, and $\chi_i$ is the set of neighbors to node $i$ and $C_i$ is the cardinality of $\chi_i$. The nodes of the WSN may be expressed as the discrete time state space sysetm with a ramp input:

$$T(n+1) = A \cdot T(n) + B \cdot t_{N+1}(n)$$
$$= A \cdot T(n) + B \cdot (\Delta T \cdot n) \quad (2)$$

where $T(n) = [t_1(n)\ t_2(n).....t_N(n)]^T$, $B$ is a column vector with binary elements 0,1 ($B = [b_1\ b_2.....b_N]^T$). If the $i$'th element of B is zero, it means that the $i$'th node is not connected to the gateway node; otherwise, there is a connection between the two. The matrix $A$ contains nonnegative elements,

$$A = [a_1\ a_2\ ..\ a_N]^T. \quad (3)$$

where the L-1 norm of $a_i$ ( $\|a_i\|_1$ ) satisfies the following:

$$\begin{bmatrix} \|a_i\|_1 = 1 & \text{if } b_i = 0 \\ \|a_i\|_1 < 1 & \text{if } b_i = 1 \end{bmatrix} \quad (4)$$

The error as a function of the discrete time can be expressed as the discrete state space system:

$$E(n+1) = A \cdot E(n) + \Delta T \cdot \hat{1} \quad (5)$$

where 
$$E(n) = \begin{bmatrix} \Delta T \cdot n - t_1(n) \\ \Delta T \cdot n - t_2(n) \\ \vdots \\ \Delta T \cdot n - t_N(n) \end{bmatrix}, \text{ and } \hat{1} = \begin{bmatrix} 1 \\ 1 \\ \vdots \\ 1 \end{bmatrix}.$$

The system in (2) is unconditionally marginally stable regardless of the connectivity of the network. It is simple to show that if the network contains a spanning path and if $B$ is not all zeros (i.e. the gateway node is connected to at least one of the WSN nodes), the system in (2) is stable with steady state error:

$$Ess = \Delta T \cdot (I - A)^{-1} \cdot \hat{1} \quad (6)$$

where $Ess = \lim_{n \to \infty} E(n)$.

C. The Stopping Criterion

Equation (6) states that the system in (2) is capable of bringing the local times of the WSN nodes arbitrarily close to that of the gateway node by simply reducing $\Delta T$. This is not practical in the case of WSN. In WSNs communication can quickly drain a node's battery. It is well- known in a WSN that the energy consumed in communicating (transmitting) one bit is equal to the energy consumed in processing one million bits within the node.

An alternative to increasing the synchronization accuracy without using excessively high communciation rates may be obtained by examining the dynamical behavior of equations (2) and (6). It is noticed that all the nodes of a network whose time upate is govered by (2) experience a strong dip in the absolute value of the error prior the steady state phase as shown in Figure-2. This is caused by the evolving local time of the node intersecting the time of the gateway node. Ideally, the error should drop to zero. However, for a finite message exachage rates, it is observed that the abosulue value of the error significantly drops for all the nodes to values well below ΔT. Formal mathematical investigation of this behavior is left for future work.

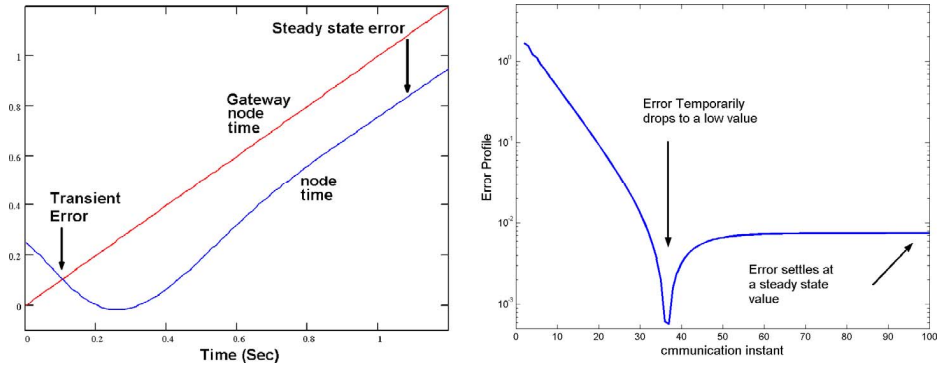

Figure-2: Time absolute erro dip in the transient phase

Currently, this observaion is investigated using intensive simulation and physical experiments. The obeservation persisted even when stochastic link connectivity was used. For example, the connectivity of the nework in Figure-3 ($t_1$ is the gateway node) is chosen as a poisson random variable:

$$f(C_{i,j}) = \begin{bmatrix} p & C_{i,j} = 1 \\ 1-p & C_{i,j} = 0 \end{bmatrix} \quad (7)$$

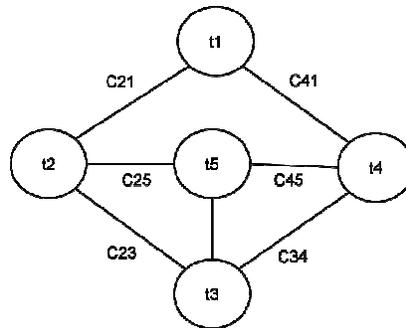

Figure-3: test nework with $t_1$ as the gateway node

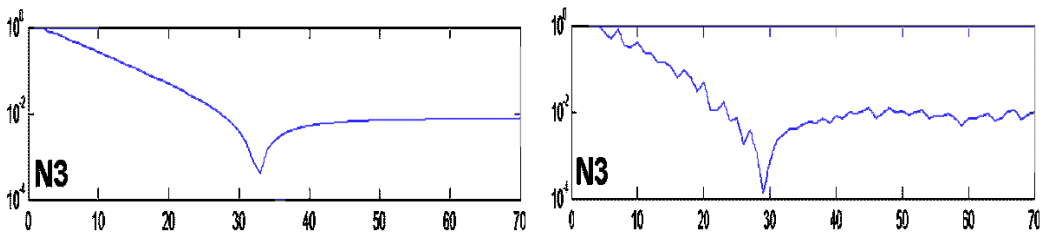

Figure-4: error profile of a node in Figure-4 with channels being on 100% of the time (p=1) and 50% of the time (p=0.5)

Figure-4 shows the error profiles of one of the nodes of the circuit for two cases (all other nodes showed similar patterns). The first case involves deterministic channels (i.e. the channels are available with probability 1, $p=1$). The other one has to do with maximum connectivity ambiguity (i.e. one has no ability to predict whether a channel is going to be available or not, $p=0.5$). As can be seen in both cases, the dip trend in the absolute value of the error persisted. Moreover, it is noticed that the times the dips occur for all the network nodes are very close. This observation provides an oppurtunity to both significantly reduce the amount of communications a node has to perform and obtain high accuracy estimates of the GWN time.

We need to have a stopping criterion (SC) to terminate the iterative process (the iterations) way before the steady state. This is carried-out by exploiting the dip behaviour which the protocol exhibit. Without this, the advantage and characteristics of the synchronization protocol will not be practical. Many SCs exist in the literature that used different stopping conditions such as maximum time, maximum number iterations, specific bound reach, mean value, standard deviation, variance, relative function, absolute function, … etc. These SCs work well in the SS region but they will not work in the transient region.

A detection filter that operates on the local time series has to be used by each node on the evolving local time estimate. This filter generates the indicator signal that will be used in localizing the instant at which the evolving local time coinside with the GWN time. A logical rule is then applied to pinpoint the local instant that will be selected as the global time. This instant is used to initialize the physical clock of the node. The rule is based on polarity change of the filter output provided that K communication cycles have elapsed. We found that K=11 works for all the cases we tested.

Although it may not be optimal, but a simple finite impulse response (FIR) difference filter to detect a change in slope can be used. This is sufficient for the scope of this work since formal design of such a filter is a mathematically involved task that requires formal investigation of the patterns the dynamics of equations (2) and (5) generate. Optimmum desing of such filter will be left for future investigation. We suggest a from of FIR difference filter with the impulse response, h($n$), given by:

$$\mathrm{h}(n) = +c_f \cdot (0.2 \cdot \delta_{n+3} + 0.5 \cdot \delta_{n+2} + 0.2 \cdot \delta_{n+1}) - 0.2 \cdot \delta_{n-1} - 0.5 \cdot \delta_{n-2} - 0.2 \cdot \delta_{n-3} \quad (8)$$

where $c_f$ is a positive constant whose value is close to unity ($0.95 \geq c_f \geq 1.05$). $c_f$ can be taken equal to 1 without affecting much the performance of the filter. This filter is found to produce good results.

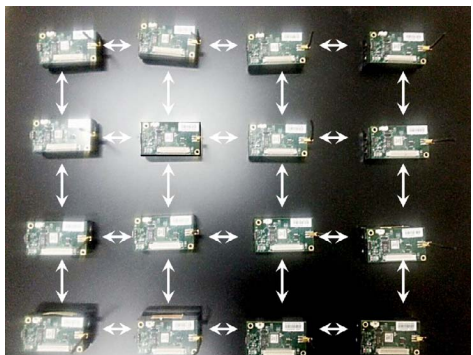
Figure-5: A sixteen-node network

### III.  RESULTS

The experiments are carried-out using the MicaZ sensor network from MEMSIC. Experiments were performed on a variety of networks. However, here we report results for a 16 node network with grid connectivity (Figure-5). The communication frequency is approximately 1KHz, Cf=1.002 and the Packet Size is 64bits. The soft clocks of the nodes were arbitrarily initialized. Figure-6 shows the evolution of the time error curves for all the nodes as a function of the server time before the network is considered synchronized and node communication seized. The data from the experiment is shown in Table-1. As can be seen, the minimum error at the peak of the dip region is considerably less than the 1 millisecond communication cycle. It is also worth noticing that the minimum error for all nodes occurred around the same time as if the network block-synchronize. The detection filter performs reasonably well yielding timing errors that are close to the minimum with number of communication cycles very close to the minimum needed. While in few cases (N12 and N15) the error is much higher than the minimum, it is still lower than the steady state error with significant saving in communication. From the above, it is obvious that better design of the detection filter can yield significant increase in synchronizing accuracy. Another reason for a better filter/decision design has to do with the noise physical implementation of the synchronizer introduces. Figure-7 shows the evolution of the error curves for a node in a four-node network along with the decision of when to stop the soft clock. While the decision filter performed well for simulation, the noise physical implementation introduced adversely affected its performance.

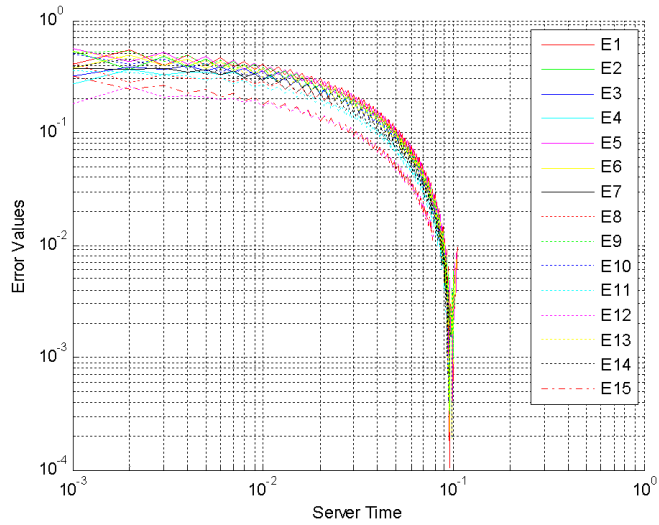

Figure-6: error of the nodes before procedure is terminated

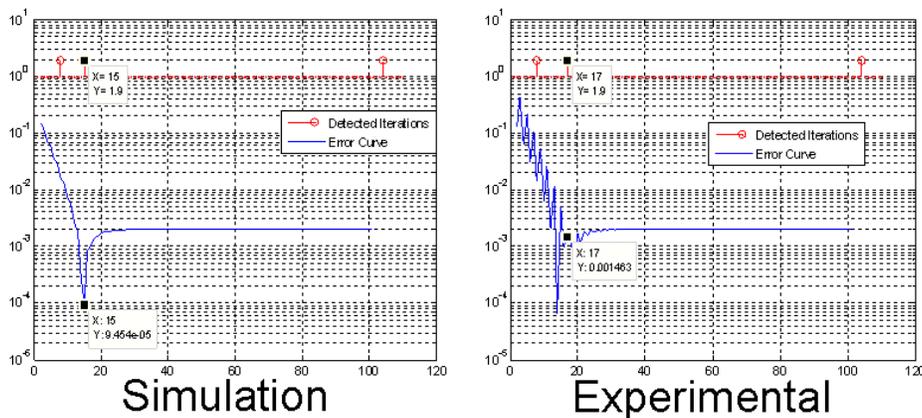

Figure-7: detection filter operating on both experimental and synthetic data

The growth in the time the suggested procedure needs for synchronization versus the size of the network is almost linear in the number of nodes (Figure-8). This indicates that the protocol is scalable for large-scale networks.

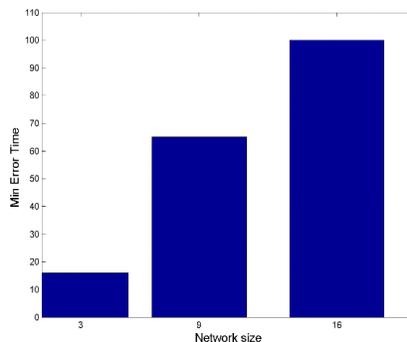

Figuer-8: Time at which minimum error occurred versus network size.

Table-1: results for the network in Figure-7

| Nodes | Min Error | | Steady State Error | | Detected Error | |
|---|---|---|---|---|---|---|
| | Comm. instant | Value | Comm. instant | Value | Comm. Instant | Value |
| N1 | 96 | 0.0001029 | 447 | 0.0413387 | 106 | 0.0095701 |
| N2 | 100 | 0.0003394 | 446 | 0.0403887 | 105 | 0.0086201 |
| N3 | 96 | 0.0003152 | 444 | 0.0381499 | 103 | 0.0028916 |
| N4 | 100 | 0.0001886 | 441 | 0.0358427 | 100 | 0.0001886 |
| N5 | 100 | 0.0003394 | 446 | 0.0403887 | 105 | 0.0086201 |
| N6 | 96 | 0.0002157 | 444 | 0.0388284 | 104 | 0.0076122 |
| N7 | 100 | 0.0002003 | 441 | 0.0353678 | 100 | 0.0002003 |
| N8 | 99 | 0.0004591 | 436 | 0.0316356 | 96 | 0.0006695 |
| N9 | 96 | 0.0003152 | 444 | 0.0381499 | 103 | 0.0028916 |
| N10 | 100 | 0.0002003 | 441 | 0.0353678 | 100 | 0.0002003 |
| N11 | 99 | 0.0002846 | 433 | 0.0290566 | 92 | 0.0024287 |
| N12 | 95 | 1.00546E-5 | 420 | 0.0208459 | 80 | 0.0140003 |
| N13 | 100 | 0.0001886 | 441 | 0.0358427 | 100 | 0.0001886 |
| N14 | 99 | 0.0004591 | 436 | 0.0316356 | 96 | 0.0006695 |
| N15 | 95 | 1.00546E-5 | 420 | 0.0208458 | 80 | 0.0140003 |
| Max | 100 | 0.0004591 | 447 | 0.0413387 | 106 | 0.0140003 |
| Min | 95 | 1.00546E-5 | 420 | 0.0208458 | 80 | 0.0001886 |

The suggested protocol is compared to the two popular protocols for WSN synchronization. The first is the Rated Flooding Time Synchronization Protocol (RFTSP) and Energy-Efficient Gradient Time Synchronization Protocol (EGTSP) [10]. RFTSP and EGTSP are compared to the suggested protocol using a nine-node WSN. Figure-9 shows the error profile of the node with maximum error value. As can be seen, the suggested protocol synchronizes faster than the other two. More importantly, it provides a clear indication of when the protocol has to stop. The above clearly show the practicality of the suggested synchronization procedure for use with WSNs in challenging environments.

## IV. CONCLUSION

The paper suggests a synchronization procedure that addresses the needs of WSNs operating in harsh environment. The experimental results and the simulation clearly demonstrate the capabilities of the suggested synchronizer. Future work will focus on mathematical analysis of the synchronizer, better design of the detection filter and experimental investigation when connectivity is stochastic.

ACKNOWLEDGMENT

The authors gratefully acknowledge the assistance of King Fahad University of Petroleum and Minerals.

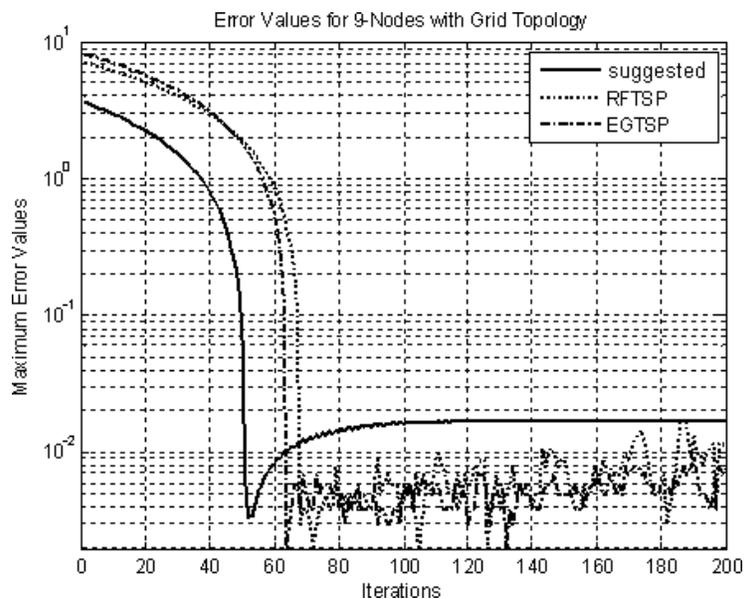

Figure-9: Maximum error curve of different protocols with 9-Grid nodes